\documentclass[twocolumn,trackchanges]{aastex701}
\usepackage{CJK}

\usepackage{amsmath}

\received{XXX}
\revised{XXX}
\accepted{XXX}

\begin{document}
\begin{CJK*}{UTF8}{gbsn}

\title{Re-interpreting the GRS 1915+105 observations within the variable TADAF scenario}

\author[orcid=0009-0009-2507-5977,sname='Xu']{Chun Xu}
\affiliation{Shanghai Astronomical Observatory, Chinese Academy of Sciences, Shanghai 200030, China}
\email[show]{chun.xuu@shao.ac.cn}

\begin{abstract}

The Galactic microquasar GRS 1915+105 is one of the most interesting X-ray binary systems, exhibiting an exceptionally diverse range of observational phenomena. It displays the most complicated and variable X-ray light curves and carries superluminal radio jets. Its X-ray and radio fluxes change on very different timescales, from seconds to minutes to days and even longer. In recent years, its jet has also displayed large position angle shifts. We find that most of the observational phenomena can be well explained within our newly developed variable turbulent ADAF (TADAF) model. The accretion disk is composed of an outer thin disk and an unstable inner TADAF torus whose size is variable. The previously suggested ``disappearance'' and regain of the inner disk is understood as the transition between the thick TADAF and the thin disk. Some types of QPOs are understood in terms of the radius change of the TADAF torus. The jet position angle shift is understood as the direct sideways frame-dragging on the jet by the Lense-Thirring effect from a tilted, spinning black hole. The slower the outshooting jet, the larger the angle shift. We note that such an effect also applies to other sources, such as V404 Cygni and the AGN M87. This mechanism may also be applicable to the famous ``precessing'' jet in SS 433.

\end{abstract}




\section{Introduction}\label{sec:intro}

GRS 1915+105 is a remarkable Galactic X-ray binary (XRB) system  discovered in 1992 \citep{CastroTirado1994}. The system consists of a stellar-mass black hole (BH) with a mass of approximately 11.2 $\pm$ 2 $M_\sun$ and a companion star of mass 0.47$\pm$0.27 $M_\sun$ at a distance of 9.4$\pm$0.8 kpc \citep{Reid2014,Reid2023,Steeghs2013}. Its orbital period is $33.85\pm0.16$ days with an inclination angle of $i=64^\circ\pm4^\circ$ \citep{Steeghs2013,Reid2023}.

This source is extraordinary for its complex and extreme variability across all electromagnetic spectrum. Observations with instruments like the Rossi X-Ray Timing Explorer (RXTE) have revealed a rich phenomenology, including dramatic X-ray flares and dips on timescales of seconds to hours to days, which are often accompanied by delayed ejections of radio-emitting plasma, and its X-ray displays a strong low frequency 0.5--6.0 Hz
quasi-periodic oscillations (QPOs) and high frequency QPO around 67 Hz \citep{Fender2004araa,Belloni2013,Morgan1997}. \citet{Belloni2000} categorized its X-ray light curves into 12 classes of distinct variable shapes, arising from rapid transitions between three fundamental states: a hard state C and two softer states A and B, thought to represent different physical configurations of the accretion disk and its inner radius. These intense accretion episodes, coupled with the ejection of relativistic jets, make GRS 1915+105 an unmatched laboratory for studying the intricate connection between accretion flow dynamics, black hole character, and the launching of powerful outflows.

GRS 1915+105 was the first Galactic source in which apparent superluminal motion ($0.92c$) was observed in its radio jets \citep{Mirabel1994}. In its ``classic'' state prior to 2019, the source launched mostly relativistic ejecta and the jets displayed a relatively stable orientation. 
A major paradigm shift occurred after 2019, when the source entered an X-ray obscured state \citep{Negoro2018,Miller2020}. High-resolution radio observations in 2025 revealed that the jets launched in this phase are fundamentally different: they are significantly slower with speeds of roughly $0.3c$ or less, and exhibit large variations in orientation, becoming misaligned with the historical continuous jet axis \citep{Jiang2026,Rodriguez2025}.

We note that most of the observed phenomena of GRS 1915+105, including its extremely strange and variable light curves, its QPOs, its episodic and continuous jet formation and their position angle (PA) shifts are not difficult to understand within the newly developed variable turbulent advection-dominated accretion flow (turbulent ADAF or TADAF) model. \citet{Xu2026jet} developed a jet formation mechanism universally applicable to active galactic nuclei (AGN), XRBs and young stellar objects (YSOs) based on a proposed turbulent ADAF torus near the central object. \citet{Xu2026xrb} proposed that the size of the turbulent ADAF is variable and used it to explain the X-ray burst phenomenon in XRBs. It unifies the presence of near-ISCO Fe emission lines with the truncated disk paradigm in the BH system like GX 339-4 and it also explains the 35-day period in the neutron star system Her X-1 through the variable TADAF sizes.
Further, \citet{Xu2026agn} explained the changing-look AGN (CLAGN) phenomenon within the variable TADAF framework and proposed a new AGN unification scheme that incorporates a turbulence-related parameter $\eta$, extending the conventional orientation-only unification paradigm. In this work, we will try to explain most of the observational facts of GRS 1915+105 under the TADAF scenario.

This paper is organized as follows. We summarize the TADAF model in Section~2. In Section~3 we discuss the observations and explanations of the X-ray light curves, QPOs, and jet PA shift in different subsections. Section~4 presents a discussion and an extension to future work. Finally, we make a short summary in Section~5.

\section{The TADAF model}

\citet{Xu2026xrb} proposed a variable TADAF model for X-ray binary systems and extended this model to AGN in \citet{Xu2026agn}. Here we summarize this model in order to explain the observations of GRS 1915+105.

The accretion flow around the central object (a BH in GRS 1915+105) consists of a standard outer thin disk \citep{Shakura1973} and an inner thick TADAF torus\citep{Xu2026xrb}. The inner part of the TADAF is inside the innermost stable circular orbit (ISCO at radius $6R_g$, where $R_g = GM/c^2$) of the black hole with mass $M$.
The boundary radius $R_T$ between the inner TADAF and the outer thin disk is not stable and may change depending on the parameter $\eta$, which stands for the turbulent energy within the TADAF expressed in units of the local Keplerian energy \citep{Xu2026jet}.
The value $\eta(R)$ generally decreases outwards from its maximum value $\eta_c$ in the innermost region. When $\eta = 0$, the TADAF transitions to a thin disk, i.e., $\eta(R_T) = 0$. Figure~1, taken from \citet{Xu2026xrb}, is an illustration of how the disk changes  with radius
under a simple sigmoid assumption for $\eta(R)$. Generally, the larger the $\eta_c$, the thicker and larger (in radius) the TADAF torus. The real situation is more complicated, though. It should be noted that the TADAF is both geometrically and optically thick \citep{Xu2026jet}, which is different from the classic ADAF of \citet{Narayan1994,Narayan1995}.

\begin{figure}
    \centering
    \includegraphics[width=0.9 \linewidth]{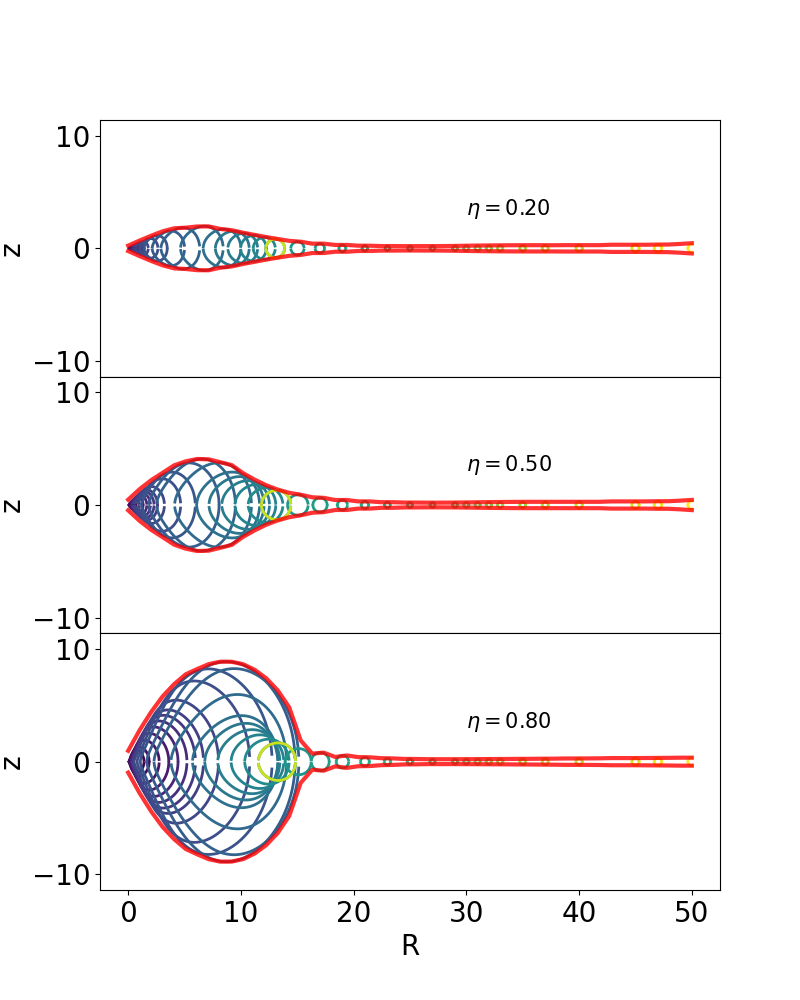}
    \caption{Shapes of the disk composed of an inner TADAF torus and an outer thin disk, for three different $\eta_c$ values indicated in the plots. Generally, the larger the $\eta_c$, the larger (in radius) and thicker the TADAF torus.}
    \label{fig:placeholder}
\end{figure}

According to \citet{Xu2026jet}, when $\eta_c > 0.5$, a jet or outflow may form. \citet{Xu2026agn} further illustrated in more detail that $\eta_c < 0.5$ represents no jet/outflow, $0.5 < \eta_c < 0.7$ represents outflow-dominated, and $\eta_c > 0.7$ jet-dominated, although these numbers are just illustrative. In the rest of the paper we mostly use $\eta$ instead of $\eta_c$ or $\eta(R)$ when no confusion is caused.

The jet energy (density) is $E_j=(2\eta-1)E_K$, where $E_K$ is the local Keplerian energy, or the jet velocity is $v_j=\sqrt{(2\eta-1)GM/R}$ in the non-relativistic approximation, where $R$ is the innermost radius of the TADAF. The half opening angle $\theta$ of the funnel in the center of the TADAF is $\sin^2\theta=1-\eta$ \citep{Xu2026jet,Frank2002}.

Now let us discuss the observed spectra, mostly following \citet{Xu2026xrb}. They are composed of three major components from different parts of the accretion disk and jet. The multicolor blackbody radiation from the outer thin disk is denoted as DBB($R_b$) \citep{Xu2026xrb,Belloni1997unstable,Shakura1973}. The power-law emission from the TADAF is denoted as PWR($\eta,R_b$) \citep{Xu2026xrb,Schlickeiser2002}. \citet{Xu2026xrb} also proposed a corona near the TADAF in order to explain some of the time-lag effects; since its spectrum is also a power law, we simply put it together with the TADAF PWR. The power-law spectrum in the jet is denoted as PWRJ($\eta$), which extends from X-ray to radio when a jet forms. The jet is formed through the collection of random bullet-like blobs ejected from the central funnel, and some radiation may also come from internal shocks within the jet. Thus the total radiation of the system is the combination of the above three types of emission:
\begin{equation}
I = DBB(R_b) + PWR(\eta,R_b) + PWRJ(\eta)
\end{equation}
All the radiation components are also related to the central mass $M$, the mass accretion rate $\dot{M}$, and other parameters such as the metallicity of the material in the accretion disk and even the magnetic field. Observational effects also depend on the inclination angle of the system and some other geometric factors. For a quickly variable system like GRS 1915+105, time-lag effects between different radiation components also need to be considered. The dynamics of the turbulence within the TADAF is virtually unknown at present and needs further study. The full radiation solution is well beyond the scope of this paper. Here we only use a semi-analytical or phenomenological analysis to compare with the observations of GRS 1915+105; we find that in most cases such a description is good enough to explain the observations qualitatively.

\section{Observations and interpretations} 

\subsection{X-ray, radio variability}

GRS 1915+105 displays extraordinary X-ray variability, from seconds to hours to days or even longer timescales. \citet{Belloni2000} found that the entire variability is made up of a limited number of variability classes, of which they identified 12. Each was observed to recur almost identically after intervals of months to even years. Later, some new classes were also found \citep{KleinWolt2002}. \citet{Belloni2000} also discovered that the complex X-ray variability can be reduced to the transition between 3 basic states: high flux soft states A and B, and low flux hard state C. The states A and B correspond to disk blackbody temperatures of $\sim 1.8$~keV and $\sim 2.2$~keV respectively, while state C corresponds to a power law with spectral index $\Gamma \sim 1.3$--$2.4$ \citep{Belloni2000,KleinWolt2002}. The high-soft state B happens mostly in the burst stage, while state A is in a relatively low flux stage. The low-hard state C is related to jet production. The low-hard (state C) and high-soft states (states B and A) are similar to those in other BH XRBs \citep{Belloni2010,Inoue2022}. The X-ray states can transit between low-hard and high-soft states very quickly within seconds, and \citet{Belloni1997unstable} suggested that the transitions from high to low states correspond to the ``disappearance'' of the inner part ($<300$~km) of the accretion disk. \citet{Fender1997} suggested that the materials were ejected as jet blobs.

It is found that the core radio flux of GRS 1915+105 also displays quick variation when it is detectable; in particular, core radio flux enhancement exactly corresponds to the low-hard state C \citep{KleinWolt2002,Mirabel1998}. That is, jet formation is directly related to the low-hard state C, as in other BH XRB systems \citep{Belloni2010,Inoue2022}.

The X-ray and radio variability in GRS 1915+105 can be understood in the variable TADAF model described in \S\,2, which was originally proposed in \citep{Xu2026xrb,Xu2026agn}. The low-hard state C corresponds to large $\eta$ and large TADAF size, while the high-soft states A and B correspond to small $\eta$ and small TADAF size. While $\eta$ vibrates, $R_T$, the boundary between the inner TADAF and the outer thin disk, changes between certain ranges as well. The X-ray flux is the combination of the DBB of the outer thin disk and the power law of the inner TADAF, with the DBB being the major component (when $R_T$ is under a certain value). Thus the vibration of $\eta$ (say between 0.3--0.8) or $R_T$ results in the X-ray flux vibration. When $\eta$ is greater than 0.5, a jet or outflow starts \citep{Xu2026jet}, resulting in radio flux enhancement. The suggestion of the ``disappearance'' of the inner disk by \citet{Belloni1997unstable} is in fact the period of TADAF growing, while their spectral fitting method can only measure the X-ray flux of the outer thin disk. This situation is the same as in GX 339-4 \citep{Xu2026xrb,Chainakun2021}. The only difference is that the transition period in GRS 1915+105 is extremely quick, within times of seconds to tens of minutes, while that of GX 339-4 is about months \citep{Chainakun2021}.

Now let us estimate the timescales of the disk transition between the thin disk and the TADAF. Since the advection velocity in the TADAF is of the order of the Keplerian velocity, the minimum transition time shall be of the order of the Keplerian period at $R_T$. For a black hole mass of $M$, the period at radius $R_T$ is $P = 2\pi \sqrt{{R_T}^3/{GM}} = 2\pi r_T^{3/2} R_g/c$, where $r_T$ is $R_T$ in units of $R_g$. For GRS 1915+105, taking a BH mass of 11 $M_\odot$, $R_T \sim 300~\mathrm{km} \sim 19~R_g$, then $P \sim 28~\mathrm{ms}$. For smaller $R_T$, the period can be even shorter. Thus the fast X-ray flux changing time of order $\sim 1~\mathrm{s}$ found by \citet{Belloni1997unstable} is actually possible. For vibrations of $R_T$ within a shorter range rather than total disappearance of TADAF, the period can be even shorter. Of course, a longer time period is also allowed, as in GX 339-4 or most other XRBs.

Careful examination of the 13 classes of X-ray light curves \citep{Belloni2000,KleinWolt2002} reveals that class $\chi$ is the most similar one to the low-hard state of other XRBs, while class $\phi$ is the most unusual type and is even difficult to understand within the TADAF model in the common sense. The X-ray flux in class $\phi$ is low but its spectrum is soft, very different from the typical low-hard and high-soft types. In the previous discussion we inadvertently took the size $R_T$ of the TADAF to be proportional to its $\eta$ value, as most classes do, or most XRBs do. However, we must consider that class $\phi$ in GRS 1915+105 is in fact a type with large $R_T$ but small $\eta$. With such a combination, its X-ray flux is likely to be low, since its DBB is weak; moreover, its PWR spectrum from the TADAF is also weak, owing to the small $\eta$---say, only around the 0.1 level---so that the turbulence in the TADAF may not be fully developed. In fact, \citet{Naik2002} found that the spectrum in class $\phi$ does not fit any of the models and requires further investigation. The many other classes are understandable within the TADAF model with the combination of states C and A, B. But the temporary variability of this source is really extreme.

\subsection{QPO and extension to QPE in AGN}
     
GRS 1915+105 displays various Quasi-Periodic Oscillations (QPOs) in X-ray (mostly $>1$~keV) light curves \citep{Fender2004,Morgan1997}. The high-frequency QPOs (e.g., 67~Hz) are often observed during the high-soft states B and A, while the low-frequency ones (e.g., 0.5--10~Hz) are mostly found in the low-hard state C. \citet{Xu2026xrb} proposed that the QPO frequencies are related to the size (radius) of the TADAF, where a larger TADAF displays low-frequency QPOs. The link between QPO frequency and state A, B, or C is consistent with that proposal. The discussion in \S\,3.1 clearly shows that some of the QPOs in GRS 1915+105 are directly linked to the vibration of the TADAF in its radius. The timescales range from less than tens of milliseconds to over tens of minutes, corresponding to a frequency range between mHz and 100~Hz, well covering the most observable QPO frequencies in GRS 1915+105.

The QPOs in infrared and radio fluxes, with typical periods of 20--40~minutes (or $\sim 0.5$~mHz), are found in GRS 1915+105 \citep{Pooley1997,Mirabel1998,Fender1998,KleinWolt2002}. They are related to the jets and outflows, which are ultimately related to the variability of the turbulent ADAF.

The many features ($>13$ classes) of the X-ray light curves in GRS 1915+105 \citep{Belloni2000,KleinWolt2002} show that the TADAF variable modes are extremely rich in this source. Some features are related to the vibration of the separating radius $R_T$ as discussed above. Other features may be related to other variable modes within the turbulent ADAF. The variable modes are temporary and alternatively switchable between different classes.

It is interesting that this vibrating TADAF QPO explanation can also be extended to explain the recently discovered Quasi-Periodic Eruptions (QPEs) phenomena found in AGNs. QPE is a class of X-ray variability phenomenon associated with intermediate-mass ($\sim 10^{5-6}\,M_\odot$) active galactic nuclei. They are characterized by intense, roughly periodic bursts of soft X-ray emission. QPEs typically exhibit recurrence timescales ranging from several hours to days, with outbursts lasting approximately one hour to several days. The X-ray spectra are well described by two components: a stable disk blackbody and a variable thermal component with characteristic temperatures of 0.1--0.2~keV, as in the source GSN 069 \citep{Miniutti2019}. \citet{Xu2026agn} interpret the QPE in GSN 069 as the vibration of the TADAF: the ``burst'' part of the blackbody X-ray emission is from the newly emerged thin disk as the TADAF shrinks inward. GRS 1915+105 provides strong support for this explanation, as its QPO shows exactly the same effect in its light curves with variable blackbody temperatures, indicating that the inner radius of the outer thin disk is vibrating (equivalently the vibration of the TADAF radius) \citep{Belloni1997unstable}. The difference in their central BH masses, i.e., tens of $M_\odot$ for BH XRBs or $\sim 10^{5-6}\,M_\odot$ for intermediate-mass AGNs, leads to the difference in their peak blackbody temperatures.

This explanation can also be extended to other QPE sources like Ansky \citep{Hernandez2025}. \citet{Guo2026} found that UV starts earlier than X-ray in a QPE eruption and diminishes later. This can be understood as follows: the shrinking of the TADAF starts at the radius $R_T$ where UV is prominent in the thin disk, and then X-ray appears when the TADAF shrinks further in, after which the TADAF expands back.

The QPE timescales in GSN 069 are the recurrence period of $\sim 9$~hours with a flare time of $\sim 1$~hour \citep{Miniutti2019}. Those in Ansky are $\sim 4.5$~days with each eruption lasting $\sim 1.5$~days in 2023, and change to $\sim 10$~days with a flare duration of about 2.5--4~days in 2024 \citep{Hernandez2025,Hernandez2025double}. Taking a BH mass of $10^6\,M_\odot$, so $R_g = 1.5\times 10^{11}$~cm, taking $R_T \sim 50\,R_g$ or $r_T=50$, then $P = 2\pi r_T^{3/2} R_g/c \approx 11100~\mathrm{s} \approx 3$~hours. Thus the observed QPE timescales in these two sources are allowed under the assumed parameters.

Under the above discussion, we tend to believe that in even more massive ($\sim 10^{8}\,M_\odot$) AGNs, QPEs may also occur but in the UV/optical wavelength rather than the X-ray/UV region, and also with different (mostly longer) timescales.

\subsection{Jet position angle shift}

The position angles (PAs) of the radio jets in GRS 1915+105 are found to change irregularly over time \citep{Dhawan2000,Rodriguez2025}. There is a mean PA direction ($147^\circ \pm 8^\circ$) and some small scattering, with a few large offsets (Fig.~3 in \citealt{Rodriguez2025}). They found that the VLA jet PA around Sep 30, 2023 is $\sim 24^\circ$ larger than the mean direction. \citet{Jiang2026} discovered both a continuous jet and discrete ejecta in the same image, with the continuous jet PA close to the mean value while the ejecta PA is in directions of $\sim 40^\circ$ and $\sim 27^\circ$ larger than the mean value, during their East Asian VLBI Network (EAVN) observations in April and September 2023, respectively (Fig.~1 in \citealt{Jiang2026}). The PA of the ejecta in September 2023 generally agrees with those of the VLA jet in \citet{Rodriguez2025} during the same period. The ejecta velocity is measured to be about $0.35c$ \citep{Jiang2026}.

The change of radio jet position angles can be understood as follows in the TADAF scenario. According to \citet{Xu2026jet}, the jet should usually be perpendicular to the accretion disk. Now suppose the black hole has spin and its spin axis lies in the accretion disk plane; then there is a Lense-Thirring (LT) effect that drags the spacetime rotating around the black hole \citep{Lense1918,Mashhoon1984}. The jet will be dragged sideways by the rotating spacetime when viewed by an outside observer. The sideways velocity is determined by the BH spin rate, so the tilt angle of the final outshoot velocity is the tangent of the ratio of the sideways velocity to the jet upward velocity. Thus, the slower the jet, the larger the tilt angle will be for the jet. According to \citet{Xu2026jet}, the jet initial velocity and funnel opening angle are directly related to the $\eta$ value. Large $\eta$ leads to a fast jet and a smaller funnel opening angle, while small $\eta$ leads to a slow jet and a wide funnel opening angle. The latter also leaves more room for the jet to tilt. The exact jet PA change is the combination of the LT effect and the funnel angle. As we have seen that the $\eta$ value in GRS 1915+105 changes dramatically within a very short time, jet/ejecta position angles can be different even within a short time period; thus we observe even 2 pairs of jet/ejecta in the same images. But the 2 pairs do occur at different moments actually.

Now let us calculate to see if the Lense-Thirring effect can make such a large-angle jet re-direction. We assume a simple situation in which a black hole of mass $M$ has angular momentum $J$, and define $a=J/M$, or the unitless $a_*=a/M=J/M^2$.
A test particle is released with vertical velocity $\beta c$ above the equator at radius $r_0$; we then calculate how much horizontal velocity the test particle will gain at a large radius.
For simplicity, we make the calculation mostly based on non-relativistic approximations.
According to \citet{Bardeen1972}, the frame-dragging angular velocity around a Kerr black hole, equivalent to the angular velocity of the zero angular momentum observer (ZAMO) rotating around the spin axis as seen from an outside reference frame, is $\omega=-g_{t\phi}/g_{\phi\phi}$, or:

\begin{equation}
    \omega(r) = \frac{2aMr}{r^4 + a^2 r^2 + 2a^2 Mr}
\end{equation}
The ZAMO velocity is $v_{\phi} = -\omega(r)r$. When the test particle moves from $r_0$ to $r_{\mathrm{out}}$ and when $r_{\mathrm{out}}$ becomes very large, its velocity gain is:
\begin{equation}
\begin{aligned}
    \Delta v_{\phi} &= \omega(r_0)\,r_0 - \omega(r_{\mathrm{out}})\,r_{\mathrm{out}} \\
    &\approx \omega(r_0)\,r_0 \\
    &=\frac{2aMr_0^2}{r_0^4 + a^2 r_0^2 + 2a^2 Mr_0} \\
    &\approx \frac{2aM}{r_0^2}   = 2a_* \left(\frac{R_g}{r_0}\right)^2 c
\end{aligned}
\end{equation}
In the final step, we restore the physical units of $c$ and $G$. This is the velocity of the test particle gained through the Lense-Thirring effect as viewed from a distant observer.

The velocity shift angle $\Delta\phi=\tan^{-1}(\Delta v_\phi /\beta c) = (2a_* / \beta )({R_g}/{r_0})^2$. Taking $a_*=1$ and assuming $r_0 = 6R_g$ at the ISCO, then $\Delta\phi$ is $3^\circ$, $6^\circ$, and $10.5^\circ$ for $\beta = 1$, $0.5$, and $0.3$, respectively. Assuming $r_0 = 3R_g$ within the ISCO, then $\Delta\phi$ is $12.5^\circ$, $24^\circ$, and $36.5^\circ$ for $\beta = 1$, $0.5$, and $0.3$, respectively. It is clear that the observed PA shift angles in the range $24^\circ$--$40^\circ$ for a velocity of $\sim 0.35c$ in GRS 1915+105 are within this LT effect caused angle shift range.

It should be noted that this calculation provides a range for the jet angle shift. When this LT shift angle is larger than the funnel opening angle $\theta$, the jet shift angle is more confined by the TADAF opening angle rather than by the LT-caused shift.
The exact observational effect of the jet velocity really depends on many parameters: the inclination angle of the accretion disk, the BH spin axis with respect to the accretion disk (we only calculate the horizontal case, not the general tilt case), the line of sight with respect to both the disk plane and the spin tilt direction, the strength of the spin $a$, and the $\eta$ value of the TADAF near the central region. Also, the jet velocity projection effect will result in changes in both the jet PA and the viewing angle (with respect to the line of sight).
One thing we may note generally is that the jet speed is directly correlated with its PA in the following way: one end of the PA corresponds to high speed while the other end corresponds to slow speed, monotonically. The jet speed and viewing angle may follow a similar monotonic relation. The real situation may be more complicated due to velocity projection and relativistic time delay effects, and the monotonic relation may not be true for all observing angle combinations. But the overall picture should be similar.

We examine Figure 3 in \citealt{Rodriguez2025} and note that at PA $174^\circ$ on September 30, 2023, the jet projected speed is 6.5~mas~day$^{-1}$ ($0.35c$) \citep{Jiang2026}; at PA $130^\circ$ on May 24, 2013, the jet projected speed is 23.6~mas~day$^{-1}$ ($0.81c$) \citep{Reid2014}; at PA $151^\circ$, near the mean PA value (year 1994/1995), the projected speed is 17.3~mas~day$^{-1}$ (Table 2 in \citealt{Reid2014}). The jet projected speeds do show a monotonic relation with the position angles, as predicted by this proposed model.

There are other XRBs that display jet PA changes.
The jet position angle and viewing angle in the XRB V404 Cygni are found to change on timescales of minutes to hours in its burst stage in 2015 \citep{Miller-Jones2019}. They explain the jet PA change as due to a precessing disk, which is caused by the Lense-Thirring effect.
Their projected speed and PA do not show a clear monotonic relationship, as examined from their Extended Data Table 3. However, it is found that of the three pairs of ejecta, their speeds and viewing angles are correlated as follows:
$\beta = 0.32 \pm 0.02, \theta = 40.6 \pm 2.4^\circ$;
$\beta = 0.35 \pm 0.01, \theta = 32.5 \pm 1.6^\circ$;
$\beta = 0.48 \pm 0.01, \theta = 14.0 \pm 0.8^\circ$ \citep{Miller-Jones2019}. They do show a monotonic relationship. Thus we believe that this proposed LT-caused jet PA shift model is suitable for the jet PA shift in V404 Cygni too.

SS 433 is a more famous XRB, likely a black hole XRB, with variable jet orientations \citep{Margon1984,Fabrika2004,Cherepashchuk2025}. The jet orientations are very well fit by the model of a precessing jet with constant velocity and a precessing period of $P_{\mathrm{prec}} = 162.375$~days. The jet precession is attributed to the accretion disk precessing, which was caused by a slaved disk due to misorientation of the companion star spin axis \citep{Whitmire1980}. However, the slaved disk model has never been firmly demonstrated. What observations display of this source are actually these two effects: the periodic change of optical emission line (H$\alpha$, H$\beta$ and more) wavelengths attributed to viewing angle change with an assumed constant jet velocity \citep{Margon1984,Fabrika2004}, and the jet PA change between $80^\circ$--$120^\circ$ in radio jets \citep{Niell1981,Paragi1999,Mioduszewski2005}. These two effects are also consequences of the proposed LT dragging effect too, although currently the precessing model gives a better precise model fitting.
The following facts are supportive of the LT dragging scenario. The X-ray jets in SS 433 do not show hollow faint and edge brightening as expected from a precessing cone; rather, they are centrally brightened with only a $20^\circ$ opening angle rather than the $40^\circ$ opening angle predicted by the precessing cone \citet{Fabrika2004}. This is more consistent with a nearly one-dimensional jet orientation moving track in stead of precessing in a cone. 
Another effect is examined in the following analysis, although it remains inconclusive.
Assume that $T_3$ is the time of the precessing phase $\psi_{\mathrm{prec}} = 0$, i.e., the time of maximum line shift or maximum disk opening (when the disk ``faces'' the observer most directly) \citep{Fabrika2004}; the precessing model then predicts that the radio jet PA at $T_3$ should lie midway between the two extreme radio jet PAs (i.e., at $\sim 100^\circ$). The LT dragging model, on the other hand, takes $T_3$ to be one end of the PA track, so that the PA is $\sim 80^\circ$ or $\sim 120^\circ$ around $T_3$.
The expected phase difference of the two models is $\Delta\psi_{\mathrm{prec}}=0.25$. 
We examined the radio jet PA in the literature, using the equation $\psi_{\mathrm{prec}} = (JD - 2443507.47) / P_{\mathrm{prec}}$, where $JD$ is the Julian Date \citep{Fabrika2004}, to calculate the jet PA at the observed epoch (phase $\psi_{\mathrm{prec}}$). We find that the measured PAs---or those estimated by eye from the radio images---are consistent with neither the precessing model nor the LT dragging model.
The estimated jet PAs at $T_3$ lie somewhere between the center of the PA range and one PA end. The result is thus inclusive. We believe such results lie in the following uncertainties: the phase shift of the precessing model over a long time was found to be about 0.1 \citep{Fabrika2004}. The radio jet and optical jet have a time shift of about 7--17~days (near 0.1 phase) \citep{Niell1981,Stirling2002}, depending on the jet length to measure its PA. The PA estimated from the images in the literature is subject to large errors. The combination of those errors is not good enough to tell the expected phase differences of 0.25. We believe that careful study of the original data with simultaneous optical line and VLBI measurements, or future dedicated optical and VLBI observations, can tell the difference.

The jets in some AGNs are found to shift too. M87 has recently been found to have a jet PA that varies with an 11-year period. \citet{Cui2023} consider the jet PA shift as due to a precessing disk caused by the LT effect, as in V404 Cygni. They did not measure the jet speed, so we cannot check the jet velocity and PA monotonic relationship. However, the jet velocity in M87 has been reported with very different values (bulk $0.27c$, $0.38c$, $\sim 0.5c$, or apparent $1$--$6c$) before \citep{Biretta1999,Walker2018,Punsly2021}; those values do not seem to be consistent with a constant velocity in a $\sim 10^\circ$ precessing model, even with velocity measurement errors. Based on the universality of the underlying physics, we believe that the proposed LT dragging scenario is also applicable to the AGN M87.

The other observational effect distinguishing the precessing model and the LT drag model lies in the following: the precessing model predicts more regular periodic precession, while the TADAF with LT drag model predicts quasi-periodic or irregular cycles. Further long-term observations will tell the difference.

\subsection{Other observational issues}
\subsubsection {Jet as ballistic ejecta combination}

\citet{Xu2026jet} proposed that the jets are a collection of the outshooting blobs from the funnel of the TADAF. The discovered ballistic ejecta and continuous jets in GRS 1915+105 are strong support for this scenario. The episodic and continuous jets both occur in this source because of the dynamic variability of the TADAF torus. The ejecta are adiabatic expansions of ballistic blobs, as discussed in \citet{Fender1997}. The continuous jets are simply more crowded blobs whose individuals are not distinguishable.
In SS 433, it is found that the typical structure of the line profiles suggests that the number of clouds (gas clumps) in the jet is $\sim 10^3$--$10^4$, and that the time to form one such cloud is $\sim 100$~s \citep{Fabrika2004}, a full support of the proposed scenario.
It is also noted that the jets in V404 Cygni are formed collectively by ballistically moving ejecta too \citep{Miller-Jones2019}.

Based on both the theoretical and observational facts, we thus believe that jets are the collective motion of ballistic blobs from the central object. It is likely that each blob has a statistically different velocity around the mean. The bright knots of the jet during propagation can be due either to intrinsically bright blobs or to internal shocks. Thus some caution is needed regarding the velocity measured from the jet blobs; sometimes the internal shocks may add additional errors to the measured velocity.

\subsubsection{Jet and Wind}
 
\citet{Neilsen2009} found that the jet and wind in GRS 1915+105 show a relatively complementary relationship, i.e., the stronger the jet, the weaker the wind, and vice versa. During the hard state, the broad Fe XXV emission line (6.7~keV) is more prominent while the jet is observed, and during the soft state, narrow Fe XXVI absorption lines (7.0~keV) with 1000~km~s$^{-1}$, which is attributed to the outflow wind, are more prominent. This phenomenon is consistent with the variable TADAF model discussed in \S\,2: the jet is prominent when $\eta > 0.7$, while the wind is prominent when $0.5 < \eta < 0.7$. It should be noted that the jet and wind are complementary but not exclusive, and the $\eta$ value of 0.7 between jet and wind is just an estimate. It is interesting to note that the mass loss from the jet and from the wind are of the same order \citep{Neilsen2009}, which is consistent with the scenario that they all originate from the central funnel. It shall be noted that, as discussed in \citet{Xu2026xrb}, if the overall accretion rate (in disk) is constant, then the outflow material is properly correlated with $\eta$. 

\subsubsection{Low flux X-ray plateau between 2018-2020}

GRS 1915+105 entered a low-flux stage in 2018 \citep{Negoro2018,Motta2021}. The X-ray flux between 2018 and 2020 is divided into 4 stages: plateau-1, flaring, plateau-2, and soft. During plateau-1 (2018), the X-ray emission remained at an unprecedentedly low level, and the radio emission was steady at a low flux density of approximately 3~mJy with limited variability. During plateau-2 (2019), which was called the X-ray obscuring stage in \citet{Motta2021}, the X-ray emission became even weaker, but the radio flux became strong and flaring (up to 100~mJy). Plateau-2 also displayed strong absorption features in X-ray spectra. Both plateaus show hard X-ray emission dominated by a power law with a spectral index of 1.91. The soft stage and pre-plateau stage are dominated by a disk blackbody spectrum.

We suggest that during both plateaus the radius $R_T$ of the TADAF is much larger than during the normal state C, resulting in a weaker X-ray flux. Moreover, $R_T$ at plateau-2 is even larger than that at plateau-1, so its X-ray flux is even weaker. The overall $\eta(R)$ during the two plateaus may be small, such as 0.1--0.2, but the central values $\eta_c$ differ between the two plateaus. During plateau-2, the central $\eta_c$ should exceed 0.5, so that the jet and winds are observable. In fact, during plateau-2, $\eta_c$ could vary rapidly between 0.3 and 0.8, making both the winds and the jet prominent. There may be a large fraction of time with $\eta_c$ around 0.4--0.6, which makes the wind dominant, so that absorption is distinctive in the spectra.
The light curves during the two plateaus are likely to be an alternation or a mixture of classes $\chi$ and $\phi$.
\citet{Motta2021} did not report thin disk radii during these plateaus; we expect that they are larger than those in the usual C state.

\section{Discussion} 

We have studied many observations of the Galactic microquasar GRS 1915+105 and try to explain them under the newly developed turbulent advection-dominated accretion flow model, and find that most of the observed phenomena are understandable in this model, at least semi-analytically. The jet structure understood as a collection of ballistic ejecta is consistent with the prediction of the jet formation mechanism proposed in \citet{Xu2026jet}, while the multi-wavelength variability of this source is consistent with the dynamic, turbulent TADAF model.

The light curves in most BH XRB systems show various characteristics.
Persistent XRB sources remain in distinct low-hard and high-soft spectral states that are associated with different timing signatures, while transient sources show outbursts with fast-rise, slow-decay profiles. The variable turbulent ADAF disk was originally proposed to explain the outbursts of transient sources \citep{Xu2026xrb}.
GRS 1915+105 displays extremely complicated X-ray light curves that other XRB sources rarely show \citep{Belloni2000,KleinWolt2002}.
They exhibit strong, aperiodic variability across a wide range of timescales, from seconds to hours to months.
We discuss in the text that such light curve variability is related to the dynamic instability of the turbulent ADAF torus. More than 13 classes of repeated light curves have been discovered in this source, and we believe there will be more types of light curves to appear in future observations. The repeated light curves may indicate certain special physical conditions of the TADAF torus. We can use such light curves to study the evolution of the TADAF, and more broadly to study the dynamics of turbulent phenomena as well.

The aperiodic fluctuations in the light curves of GRS 1915+105 often display a power-law-like power density spectrum and are accompanied by QPOs at frequencies ranging from mHz to kHz \citep{Fender2004,Morgan1997}.
As discussed in the text, some types of low-frequency QPOs in GRS 1915+105 are related to the vibration of the TADAF in its radius. This is clearly seen in the blackbody temperature changes during the state transitions \citep{Belloni1997unstable}.
It is interesting to investigate how correspondingly the power-law radiation spectrum changes along with the blackbody temperature change. The power-law spectrum, as discussed in the text, is a distinctive indication of the TADAF.
Setting the TADAF as the base model, we can study its dynamics using QPO observations to see how many different modes will appear in the accretion flow, and how they are related to the later evolution of the TADAF, jet, outflow, and accretion.

It is not an overemphasis of its importance that GRS 1915+105 provides with its light curve samples. While many other XRBs show typical large variations of their X-ray flux on timescales of months, e.g., GX 339-4 \citep{Chainakun2021}, GRS 1915+105 displays large-amplitude light curves on timescales from months down to minutes or even seconds. As this is related to the large-scale switch between the inner torus and the disk, it is really amazing to see that a torus/disk with a size of hundreds of kilometers can switch within seconds.

The blackbody temperature change in GRS 1915+105 within the state transition \citep{Belloni1997unstable} is a clear indication of the change of the TADAF radius. A similar temperature change phenomenon was found in the QPE sources, e.g., GSN 069 and Ansky, suggesting that the same mechanism is at work in these intermediate-massive AGNs \citep{Xu2026agn}. Since the radius change of the TADAF seems to be common due to its turbulent nature, we expect that QPE-like phenomena can be searched for at UV/optical wavelengths in more massive AGNs, but on longer timescales.

The change of jet position angle is not uncommon in XRB systems. It is usually explained with the precessing disk model \citep{Margon1984,Katz1973,Jiang2026}. The discovery of irregular jet PA changes in GRS 1915+105 \citep{Rodriguez2025}, and both the continuous jet and separate ejecta that move in different directions simultaneously in one radio image \citep{Jiang2026}, is not fully consistent with the precessing disk model, which expects more or less periodic jet PA changes.
We proposed a model in which the black hole spin axis lies in the plane of the accretion disk while the jet is originally perpendicular to the disk, and the Lense-Thirring effect drags the jet offside. We make a simple calculation and find that this scenario, along with the variable $\eta$ and corresponding funnel opening angle effect, is enough to account for the jet PA shift. The simultaneous continuous jet and ejecta appearing in one image is in fact because they occur at different times, but the quick variability of the TADAF disk makes them appear in the same image. The continuous jets are usually fast while the ejecta are slow, corresponding to different $\eta$ values at the ejection. It appears that the jet that is faster is composed of more ballistic ejecta together, so it appears continuous, while the separate ejecta, which are slow, are composed of fewer ejecta, so they appear separate. But such ''more or less" is relative and may not be an intrinsic characteristic.

The proposed scenario of an LT-drag-induced jet PA shift predicts several observational effects: the jet velocity is faster at one PA end and slow at the other PA end. Due to the velocity projection effect, the PA can also be related to the viewing angle. Such an effect is very clear in GRS 1915+105, and generally consistent in another XRB, V404 Cygni. For SS 433, the situation is more complicated since the velocity is measured/calculated through the precessing disk and based on the optical radial projection velocity. We suggest that the radio jet PA at $T_3$ shall lie in the middle of the two PA ends based on the precessing model, while the jet PA shall lie at one end of the PA based on the LT drag model. The result we found in the literature is inconclusive. More careful analysis or new observations are needed to distinguish these two models. But based on a more uniform physical picture, we believe that the jet PA shift in SS 433 is also caused by the proposed LT drag effect. For the AGN M87, the jet is also found to shift with a tentative 11-year period. Since no corresponding jet velocity was measured, we suggest more dedicated observations need to be performed on this source.

If this Lense-Thirring drag jet PA scenario is confirmed, then we believe that we can use this effect to study the black hole spin. Our calculation is very simple. For a detailed quantitative comparison, a full relativity calculation with the gravitomagnetic force formula for a tilted black hole spin is needed.

The radiation spectrum from the TADAF is expected to be a power law, as first proposed by \citet{Xu2026xrb} based on the Kolmogorov power spectrum of turbulence and the second-order Fermi acceleration mechanism \citep{Schlickeiser2002}. The typical spectral index is 2. However, how the radiation spectrum evolves with $\eta$ has not been fully studied. In particular, when a light curve class such as class $\phi$ in GRS 1915+105 is observed with supposedly small $\eta$ values, its spectrum cannot yet be well modeled. Thus, the study of the TADAF spectral characteristics as a function of the $\eta$ parameter is an important avenue for future work.

\section{Summary}

The Galactic microquasar GRS 1915+105 displays extraordinarily rich observational phenomena in its X-ray, infrared, and radio light curves and in its jet velocity and orientations. We interpret many of the observations under the newly developed TADAF model and find that most of the observations are consistent with this model. Here is the summary:

(1) The many classes of light curves are the result of TADAF torus vibrations. In particular, the early interpretation of the ``disappearance'' and regain of the inner disk is now interpreted as the transition between the thick TADAF and thin disk.

(2) Some types of QPOs are the radius vibration of the TADAF. This variable TADAF radius can also be extended to explain the QPE phenomenon in intermediate-massive AGNs.

(3) The observed variable jet velocity, ranging from superluminal $0.92c$ to $0.35c$, is governed by the $\eta$ parameter of the TADAF.

(4) The change in jet PA is caused by the Lense-Thirring drag effect of a tilted, spinning black hole. This scenario can be extended to explain the jet PA changes observed in other XRB systems---V404 Cygni and SS 433---and even in the AGN M87.

(5) The jet is formed through collections of ballistic ejecta of the normal baryonic material from the accretion disk; the continuous jet and episodic jet are simply due to different crowdedness of the ejecta, and the on-set/off-set of the jet is due to the variable $\eta$ values.

(6) The jet and outflow winds are complementary; they all originate from the central funnel, and different $\eta$ values determine which component is dominant.

\begin{acknowledgments}

The author is thankful to Dr. Hengxiao Guo for providing some information on the source Ansky. This work is supported by the China Manned Space Program Grant No. CMS-CSST-2025-A19 and grants allocated for the development of the Multi-Channel Imager and the Integral Field Spectrograph, led by Drs. Zhenya Zheng and Lei Hao, respectively, for the Chinese Space-station Survey Telescope (CSST).

\end{acknowledgments}

\bibliography{sample701}{}

@article{Xu2026jet,
  author = {Xu, Chun},
  title = {A jet formation model for astrophysical objects},
  journal = {\mnras},
  year = {2026},
  volume = {550},
  number = {2},
  pages = {1--8},
  month = {August},
  doi = {10.1093/mnras/stag1236},
  publisher = {Oxford University Press (OUP)},
  key = {Xu2026a}
}

@article{Xu2026xrb,
    author    = {Xu, Chun},
    title     = {A variable {ADAF} disk model for {X}-ray binary systems}, 
    journal   = {arXiv e-prints},
    year      = {2026},
    month     = {March},
    adsurl    = {https://arxiv.org/abs/2603.10311},
    archivePrefix = {arXiv},
    eprint    = {2603.10311},
    primaryClass = {astro-ph.HE},
    key = {Xu2026b}
}

@article{Xu2026agn,
  author = {Xu, Chun},
  title = {Variable ADAF disk as the origin of Changing-Look AGN},
  journal = {arXiv e-prints},
  year = {2026},
  month = {March},
  adsurl    = {https://arxiv.org/abs/2603.28666},
  archivePrefix = {arXiv},
  eprint = {2603.28666},
  primaryClass = {astro-ph.HE},
  note = {Preprint},
  key = {Xu2026c}
}

@book{Frank2002,
    author = {Frank, J. and King, A. and Raine, D. J.},
    title = {Accretion Power in Astrophysics},
    publisher = {Cambridge University Press},
    address = {Cambridge, UK},
    year = {2002},
    isbn = {978-0521629577},
    doi = {10.1017/CBO9781139164245}
}

@article{KleinWolt2002,
  author = {Klein-Wolt, M. and Fender, R. P. and Pooley, G. G. and Belloni, T. and Migliari, S. and Morgan, E. H. and van der Klis, M.},
  title = {Hard X-ray states and radio emission in GRS 1915+105},
  journal = {\mnras},
  volume = {331},
  number = {3},
  pages = {745--764},
  year = {2002},
  month = {April},
  doi = {10.1046/j.1365-8711.2002.05223.x},
  publisher = {Oxford University Press (OUP)}
}

@article{Jiang2026,
  author = {Jiang, Wu and Yan, Xi and Yan, Zhen and Li, Ya-Ping and Cui, Lang and Shen, Zhi-Qiang},
  title = {A Large Misalignment between Continuous Jet and Discrete Ejecta in Microquasar GRS 1915+105 during Its Obscured Phase},
  journal = {\apjl},
  volume = {1000},
  number = {2},
  pages = {L45},
  year = {2026},
  month = {3},
  doi = {10.3847/2041-8213/ae5186},
  publisher = {American Astronomical Society}
}

@article{Rodriguez2025,
  author = {Rodríguez, Luis F. and Mirabel, I. Félix},
  title = {An Unusual Change in the Radio Jets of GRS 1915+105},
  journal = {\apj},
  volume = {986},
  number = {1},
  pages = {108},
  year = {2025},
  month = {June},
  doi = {10.3847/1538-4357/adda33},
  publisher = {American Astronomical Society},
  adsurl = {https://ui.adsabs.harvard.edu/abs/2025ApJ...986..108R}
}

@article{Belloni2000,
  author = {Belloni, T. and Klein-Wolt, M. and Mendez, M. and van der Klis, M. and van Paradijs, J.},
  title = {A model-independent analysis of the variability of GRS 1915+105},
  journal = {A\&A},
  volume = {355},
  pages = {271--290},
  year = {2000},
  doi = {https://doi.org/10.48550/arXiv.astro-ph/0001103}
}

@article{Mirabel1994,
  author = {Mirabel, I. F. and Rodríguez, L. F.},
  title = {A superluminal source in the Galaxy},
  journal = {Nature},
  volume = {371},
  number = {6492},
  pages = {46--48},
  year = {1994},
  doi = {10.1038/371046a0}
}

@article{CastroTirado1994,
  author = {Castro-Tirado, Alberto J. and Brandt, Soren and Lund, Niels and Lapshov, Igor and Sunyaev, Rashid A. and Shlyapnikov, Aleksei A. and Guziy, Sergei and Pavlenko, Elena P.},
  title = {Discovery and observations by watch of the X-ray transient GRS 1915+105},
  journal = {ApJS},
  volume = {92},
  pages = {469},
  year = {1994},
  doi = {10.1086/191998}
}

@article{Inoue2022,
    author = {Inoue, H.},
    title = {X-ray observations of accretion disks},
    journal = {\pasj},
    volume = {74},
    pages = {1},
    year = {2022},
    doi = {10.1093/pasj/psab066}
}

@incollection{Belloni2010,
    author = {Belloni, T. M.},
    title = {States and Transitions in Black Hole Binaries},
    editor = {Belloni, T.},
    booktitle = {The Jet Paradigm},
    series = {Lecture Notes in Physics},
    volume = {794},
    publisher = {Springer},
    address = {Berlin},
    year = {2010},
    pages = {1},
    doi = {10.1007/978-3-540-76937-8_1}
}

@article{Fender2004,
    author = {Fender, R. P. and Belloni, T. M. and Gallo, E.},
    title = {Towards a unified model for black hole X-ray binary jets},
    journal = {MNRAS},
    volume = {355},
    pages = {1105--1118},
    year = {2004},
    doi = {10.1111/j.1365-2966.2004.08384.x}
}

@article{Shakura1973,
    author = {Shakura, N. I. and Sunyaev, R. A.},
    title = {Black holes in binary systems: Observational appearance},
    journal = {\aap},
    volume = {24},
    pages = {337--355},
    year = {1973},
    adsurl = {https://ui.adsabs.harvard.edu/abs/1973A&A....24..337S}
}

@article{Narayan1994,
    author = {Narayan, R. and Yi, I.},
    title = {Advection-dominated accretion: A self-similar solution},
    journal = {ApJ},
    volume = {428},
    pages = {L13--L16},
    year = {1994},
    doi = {10.1086/187381}
}

@article{Narayan1995,
    author = {Narayan, R. and Yi, I.},
    title = {Advection-dominated accretion: Underfed black holes and neutron stars},
    journal = {ApJ},
    volume = {452},
    pages = {710--735},
    year = {1995},
    doi = {10.1086/176343}
}

@book{Schlickeiser2002,
    author = {Schlickeiser, R.},
    title = {Cosmic Ray Astrophysics},
    series = {Astronomy and Astrophysics Library},
    publisher = {Springer},
    address = {Berlin, Heidelberg},
    year = {2002},
    isbn = {978-3-662-04814-6},
    doi = {10.1007/978-3-662-04814-6}
}

@article{Belloni1997unstable,
  title = {An Unstable Central Disk in the Superluminal Black Hole X-Ray Binary GRS 1915+105},
  volume = {479},
  issn = {0004-637X},
  doi = {10.1086/310595},
  number = {2},
  journal = {\apj},
  publisher = {American Astronomical Society},
  author = {Belloni, T. and M{\'{e}}ndez, M. and King, A. R. and van der Klis, M. and van Paradijs, J.},
  year = {1997},
  month = apr,
  pages = {L145–L148}
}

@article{Miniutti2019,
  title = {Nine-hour X-ray quasi-periodic eruptions from a low-mass black hole galactic nucleus},
  volume = {573},
  url = {https://www.nature.com/articles/s41586-019-1556-x},
  doi = {10.1038/s41586-019-1556-x},
  number = {7774},
  journal = {Nature},
  publisher = {Springer Science and Business Media {LLC}},
  author = {Miniutti, G. and Saxton, R. D. and Giustini, M. and Alexander, K. D. and Fender, R. P. and Heywood, I. and Monageng, I. and Coriat, M. and Tzioumis, A. K. and Read, A. M. and Knigge, C. and Gandhi, P. and Pretorius, M. L. and Ag{\'{i}}s-Gonz{\'{a}}lez, B.},
  year = {2019},
  month = sep,
  pages = {381–384}
}

@article{Fender1997,
  author = {Fender, R. P. and Pooley, G. G. and Brocksopp, C. and Newell, S. J.},
  title = {Rapid infrared flares in GRS 1915+105: evidence for infrared synchrotron emission},
  journal = {MNRAS},
  volume = {290},
  number = {4},
  pages = {L65--L69},
  year = {1997},
  month = {October},
  doi = {10.1093/mnras/290.4.L65},
  eprint = {astro-ph/9707317},
  primaryClass = {astro-ph}
}

@article{Fender1998,
  author = {Fender, R. P. and Pooley, G. G.},
  title = {Infrared synchrotron oscillations in GRS 1915+105},
  journal = {MNRAS},
  volume = {300},
  number = {2},
  pages = {573--576},
  year = {1998},
  month = {October},
  doi = {10.1046/j.1365-8711.1998.01921.x},
  eprint = {astro-ph/9806073},
  primaryClass = {astro-ph}
}

@article{Mirabel1998,
  author = {Mirabel, I. F. and Dhawan, V. and Chaty, S. and Rodr{\'i}guez, L. F. and Mart{\'i}, J. and Robinson, C. R. and Swank, J. and Geballe, T. R.},
  title = {Accretion instabilities and jet formation in GRS 1915+105},
  journal = {A\&A},
  volume = {330},
  pages = {L9--L12},
  year = {1998},
  month = {feb},
  doi = {https://doi.org/10.48550/arXiv.astro-ph/9711097},
  eprint = {astro-ph/9711097},
  primaryClass = {astro-ph}
}

@article{Chainakun2021,
    author = {Chainakun, P. and Ertan, {\"U}.},
    title = {Quasi-periodic oscillations in neutron star X-ray binaries},
    journal = {\aap},
    volume = {645},
    pages = {A99},
    year = {2021},
    doi = {10.1051/0004-6361/202039090}
}

@article{Morgan1997,
  author = {Morgan, E. H. and Remillard, R. A. and Greiner, J.},
  title = {RXTE Observations of QPOs in the Black Hole Candidate GRS 1915+105},
  journal = {ApJ},
  volume = {482},
  number = {2},
  pages = {993--1010},
  year = {1997},
  month = {June},
  doi = {10.1086/304191}
}

@article{Hernandez2025,
  author = {Hern\'andez-Garc\'ia, Lorena and Chakraborty, Joheen and S\'anchez-S\'aez, Paula and Ricci, Claudio and Cuadra, Jorge and McKernan, Barry and Ford, K. E. Saavik and Ar\'evalo, Patricia and Rau, Arne and Arcodia, Riccardo and Kara, Erin and Liu, Zhu and Merloni, Andrea and Bruni, Gabriele and Goodwin, Adelle and et al.},
  title = {Discovery of extreme quasi-periodic eruptions in a newly accreting massive black hole},
  journal = {Nature Astronomy},
  volume = {9},
  number = {6},
  pages = {895--906},
  year = {2025},
  month = {April},
  doi = {10.1038/s41550-025-02523-9},
  publisher = {Springer Science and Business Media LLC}
}

@article{Guo2026,
  author = {Guo, Hengxiao and Yan, Zhen and Li, Ya-Ping and Chakraborty, Joheen and S\'anchez-S\'aez, Paula and Hern\'andez-Garc\'ia, Lorena and Zhang, Wenda and Sun, Jingbo and Li, Shuang-Liang and Deng, Hongping},
  title = {Evidence for a Delayed Ultraviolet Counterpart to X-Ray Quasiperiodic Eruptions in Ansky},
  journal = {ApJL},
  volume = {1000},
  number = {2},
  pages = {L57},
  year = {2026},
  month = {March},
  doi = {10.3847/2041-8213/ae524b}
}

@article{Reid2014,
       author = {{Reid}, M.~J. and {McClintock}, J.~E. and {Steiner}, J.~F. and {Steeghs}, D. and {Remillard}, R.~A. and {Dhawan}, V. and {Narayan}, R.},
        title = "{A Parallax Distance to the Microquasar GRS 1915+105 and a Revised Estimate of its Black Hole Mass}",
      journal = {\apj},
         year = {2014},
        month = {nov},
       volume = {796},
       number = {1},
        pages = {2},
          doi = {10.1088/0004-637X/796/1/2},
archivePrefix = {arXiv},
       eprint = {1406.0018},
}

@article{Mashhoon1984,
  author = {Mashhoon, B. and Hehl, F. W. and Theiss, D. S.},
  title = {On the gravitational effects of rotating masses: The Thirring-Lense papers},
  journal = {General Relativity and Gravitation},
  volume = {16},
  number = {8},
  pages = {711--750},
  year = {1984},
  doi = {10.1007/BF00762913}
}

@article{Lense1918,
  author = {Lense, J. and Thirring, H.},
  title = {\"Uber den Einflu\ss\ der Eigenrotation der Zentralk\"orper auf die Bewegung der Planeten und Monde nach der Einsteinschen Gravitationstheorie},
  journal = {Phys. Z.},
  volume = {19},
  pages = {156--163},
  year = {1918}
}

@article{Miller-Jones2019,
       author = {{Miller-Jones}, James C. A. and {Tetarenko}, Alexandra J. and
                 {Sivakoff}, Gregory R. and {Middleton}, Matthew J. and
                 {Altamirano}, Diego and {Anderson}, Gemma E. and {Belloni}, Tomaso M. and
                 {Fender}, Rob P. and {Jonker}, Peter G. and {K{\"o}rding}, Elmar G. and
                 {Krimm}, Hans A. and {Maitra}, Dipankar and {Markoff}, Sera and
                 {Migliari}, Simone and {Mooley}, Kunal P. and {Rupen}, Michael P. and
                 {Russell}, David M. and {Russell}, Thomas D. and {Sarazin}, Craig L. and
                 {Soria}, Roberto and {Tudose}, Valeriu},
        title = "{A rapidly changing jet orientation in the stellar-mass black hole system V404 Cygni}",
      journal = {\nat},
         year = {2019},
        month = {may},
       volume = {569},
       number = {7756},
        pages = {374--377},
          doi = {10.1038/s41586-019-1152-0},
archivePrefix = {arXiv},
       eprint = {1906.05400},
}

@article{Cui2023,
  author = {Cui, Yuzhu and Hada, Kazuhiro and Kawashima, Tomohisa and Kino, Motoki and Lin, Weikang and Mizuno, Yosuke and others},
  title = {Precessing jet nozzle connecting to a spinning black hole in M87},
  journal = {Nature},
  volume = {621},
  number = {7980},
  pages = {711--715},
  year = {2023},
  doi = {10.1038/s41586-023-06479-6}
}

@article{Fabrika2004,
  author = {Fabrika, S.},
  title = {The jets and supercritical accretion disk in SS433},
  journal = {Astrophysics and Space Physics Reviews},
  volume = {12},
  pages = {1--152},
  year = {2004},
  doi = {10.48550/arXiv.astro-ph/0603390},
  eprint = {astro-ph/0603390},
  primaryClass = {astro-ph}
}

@article{Margon1984,
  author = {Margon, Bruce},
  title = {Observations of SS 433},
  journal = {ARA\&A},
  volume = {22},
  pages = {507--536},
  year = {1984},
  doi = {10.1146/annurev.aa.22.090184.002451}
}

@article{Cherepashchuk2025,
  author = {Cherepashchuk, A. M. and Dodin, A. V. and Postnov, K. A.},
  title = {Unique microquasar SS433: new results, new issues},
  journal = {Phys. Usp.},
  volume = {68},
  number = {10},
  pages = {1042--1060},
  year = {2025},
  doi = {10.3367/UFNe.2025.05.039904},
  eprint = {2506.01106},
  primaryClass = {astro-ph.HE}
}

@article{Neilsen2009,
  author  = {Neilsen, Joseph and Lee, Julia C.},
  title   = {Accretion disk winds as the jet suppression mechanism in the microquasar {GRS} 1915+105},
  journal = {Nature},
  year    = {2009},
  volume  = {458},
  number  = {7237},
  pages   = {481--484},
  doi     = {10.1038/nature07680},
  issn    = {0028-0836},
}

@article{Negoro2018,
  author  = {Negoro, H. and Kawai, N. and Nakajima, M. and Mihara, T. and Sugizaki, M. and Ueno, S. and Tomida, H. and Serino, M. and Ishikawa, M. and Nakahira, S. and others},
  title   = {{MAXI/GSC} detection of an unusually low X-ray state of {GRS} 1915+105},
  journal = {The Astronomer's Telegram},
  year    = {2018},
  volume  = {11828},
  pages   = {1},
  month   = {jul},
}

@article{Motta2021,
  author  = {Motta, S. E. and Kajava, J. J. E. and Giustini, M. and Williams, D. R. A. and Del Santo, M. and Fender, R. and Green, D. A. and Heywood, I. and Rhodes, L. and Segreto, A. and Sivakoff, G. and Woudt, P. A.},
  title   = {Observations of a radio-bright, X-ray obscured GRS 1915+105},
  journal = {\mnras},
  year    = {2021},
  volume  = {503},
  number  = {1},
  pages   = {152--161},
  doi     = {10.1093/mnras/stab511},
  eprint  = {2101.01187},
  archivePrefix = {arXiv},
  primaryClass  = {astro-ph.HE},
}

@article{Bardeen1972,
  author  = {Bardeen, James M. and Press, William H. and Teukolsky, Saul A.},
  title   = {Rotating Black Holes: Locally Nonrotating Frames, Energy Extraction, and Scalar Synchrotron Radiation},
  journal = {\apj},
  year    = {1972},
  volume  = {178},
  pages   = {347--370},
  doi     = {10.1086/151796},
  adsurl  = {https://ui.adsabs.harvard.edu/abs/1972ApJ...178..347B},
}

@article{Naik2002,
  author  = {Naik, S. and Rao, A. R. and Chakrabarti, Sandip K.},
  title   = {Fast Transition between High-soft and Low-soft States in {GRS} 1915+105: Evidence for a Critically Viscous Accretion Flow},
  journal = {Journal of Astrophysics and Astronomy},
  year    = {2002},
  volume  = {23},
  pages   = {213},
  doi     = {10.1007/BF02702284},
  eprint  = {astro-ph/0211515},
  archivePrefix = {arXiv},
  primaryClass  = {astro-ph},
}

@article{Reid2023,
   author = {{Reid}, M.~J. and {Miller-Jones}, J.~C.~A.},
    title = "{On the Distances to the X-Ray Binaries Cygnus X-3 and GRS 1915+105}",
  journal = {\apj},
     year = {2023},
    month = {dec},
   volume = {959},
   number = {2},
    pages = {85},
      doi = {10.3847/1538-4357/acfe0c},
   adsurl = {https://ui.adsabs.harvard.edu/abs/2023ApJ...959...85R}
}

@article{Steeghs2013,
   author = {{Steeghs}, D. and {McClintock}, J.~E. and {Parsons}, S.~G. and {Reid}, M.~J. and {Littlefair}, S. and {Dhillon}, V.~S.},
    title = "{The Not-so-massive Black Hole in the Microquasar GRS1915+105}",
  journal = {\apj},
     year = {2013},
    month = {apr},
   volume = {768},
   number = {2},
    pages = {185},
      doi = {10.1088/0004-637X/768/2/185},
   adsurl = {https://ui.adsabs.harvard.edu/abs/2013ApJ...768..185S}
}

@article{Fender2004araa,
   author = {{Fender}, R.~P. and {Belloni}, T.},
    title = "{GRS 1915+105 and the Disc-Jet Coupling in Accreting Black Hole Systems}",
  journal = {\araa},
     year = {2004},
    month = {sep},
   volume = {42},
   number = {1},
    pages = {317-364},
      doi = {10.1146/annurev.astro.42.053102.134031},
   adsurl = {https://ui.adsabs.harvard.edu/abs/2004ARA&A..42..317F}
}

@article{Belloni2013,
   author = {{Belloni}, T.~M. and {Altamirano}, D.},
    title = "{High-frequency quasi-periodic oscillations from GRS 1915+105}",
  journal = {\mnras},
     year = {2013},
    month = {jun},
   volume = {432},
   number = {1},
    pages = {10-18},
      doi = {10.1093/mnras/stt500},
   adsurl = {https://ui.adsabs.harvard.edu/abs/2013MNRAS.432...10B}
}

@artile{Miller2020,
   author = {{Miller}, J.~M. and {Zoghbi}, A. and {Raymond}, J. and {Balakrishnan}, M. and {Brenneman}, L. and {Cackett}, E. and {Draghis}, P. and {Fabian}, A.~C. and {Gallo}, E. and {Kaastra}, J. and {Kallman}, T. and {Kammoun}, E. and {Motta}, S.~E. and {Proga}, D. and {Reynolds}, M.~T. and {Trueba}, N.},
    title = "{An Obscured, Seyfert 2-like State of the Stellar-mass Black Hole GRS 1915+105 Caused by Failed Disk Winds}",
  journal = {\apj},
     year = {2020},
    month = {nov},
   volume = {904},
   number = {1},
    pages = {30},
      doi = {10.3847/1538-4357/abbb31},
   adsurl = {https://ui.adsabs.harvard.edu/abs/2020ApJ...904...30M}
}

@article{Pooley1997,
   author = {{Pooley}, G.~G. and {Fender}, R.~P.},
    title = "{The variable radio emission from GRS 1915+105}",
  journal = {\mnras},
     year = 1997,
    month = dec,
   volume = {292},
   number = {4},
    pages = {925-933},
      doi = {10.1093/mnras/292.4.925},
   adsurl = {https://ui.adsabs.harvard.edu/abs/1997MNRAS.292..925P}
}

@article{Dhawan2000,
   author = {{Dhawan}, V. and {Mirabel}, I.~F. and {Rodr{\'i}guez}, L.~F.},
    title = "{AU-Scale Synchrotron Jets and Superluminal Ejecta in GRS 1915+105}",
  journal = {\apj},
     year = {2000},
    month = {nov},
   volume = {543},
   number = {1},
    pages = {373-385},
      doi = {10.1086/317088},
   adsurl = {https://ui.adsabs.harvard.edu/abs/2000ApJ...543..373D}
}

@article{Hernandez2025double,
   author = {{Hern{\'a}ndez-Garc{\'i}a}, L. and {S{\'a}nchez-S{\'a}ez}, P. and {Chakraborty}, J. and {Cuadra}, J. and {Miniutti}, G. and {Arcodia}, R. and {Ar{\'e}valo}, P. and {Giustini}, M. and {Kara}, E. and {Ricci}, C. and {Pasham}, D.~R. and {Arzoumanian}, Z. and {Gendreau}, K. and {Lira}, P.},
    title = "{NICER observations reveal doubled timescales in Ansky's quasi-periodic eruptions}",
  journal = {\aap},
     year = {2025},
    month = {nov},
   volume = {703},
    pages = {A263},
      doi = {10.1051/0004-6361/202555258},
   adsurl = {https://ui.adsabs.harvard.edu/abs/2025A&A...703A.263H}
}

@article{Katz1973,
   author = {{Katz}, J.~I.},
    title = "{Thirty-five-day Periodicity in Her X-1}",
  journal = {Nature Physical Science},
     year = {1973},
    month = {nov},
   volume = {246},
   number = {155},
    pages = {87-89},
      doi = {10.1038/physci246087a0},
   adsurl = {https://ui.adsabs.harvard.edu/abs/1973NPhS..246...87K}
}

@article{Whitmire1980,
       author = {{Whitmire}, D.~P. and {Matese}, J.~J.},
        title = "{The slaved disc model for SS 433.}",
      journal = {\mnras},
         year = {1980},
        month = {dec},
       volume = {193},
        pages = {707-712},
          doi = {10.1093/mnras/193.4.707},
       adsurl = {https://ui.adsabs.harvard.edu/abs/1980MNRAS.193..707W}
}

@article{Niell1981,
       author = {{Niell}, A.~E. and {Lockhart}, T.~G. and {Preston}, R.~A.},
        title = "{Periodic changes in the compact radio structure of SS 433.}",
      journal = {\apj},
         year = {1981},
        month = {nov},
       volume = {250},
        pages = {248-253},
          doi = {10.1086/159369},
       adsurl = {https://ui.adsabs.harvard.edu/abs/1981ApJ...250..248N}
}

@article{Paragi1999,
       author = {{Paragi}, Z. and {Vermeulen}, R.~C. and {Fejes}, I. and {Schilizzi}, R.~T. and {Spencer}, R.~E. and {Stirling}, A.~M.},
        title = "{The inner radio jet region and the complex environment of SS 433.}",
      journal = {\aap},
         year = {1999},
        month = {aug},
       volume = {348},
        pages = {910-916},
archivePrefix = {arXiv},
       eprint = {astro-ph/9907169},
       adsurl = {https://ui.adsabs.harvard.edu/abs/1999A&A...348..910P}
}

@inproceedings{Mioduszewski2005,
       author = {{Mioduszewski}, A.~J. and {Dhawan}, V. and {Rupen}, M.~P.},
        title = "{Recent Results of VLBA Imaging of X-Ray Binaries: the Newest and Oldest Microquasars}",
    booktitle = {Future Directions in High Resolution Astronomy},
         year = {2005},
       editor = {{Romney}, J.~D. and {Reid}, M.~J.},
       series = {ASPC Series},
       volume = {340},
        pages = {281},
       adsurl = {https://ui.adsabs.harvard.edu/abs/2005ASPC..340..281M}
}

@article{Stirling2002,
       author = {{Stirling}, A.~M. and {Jowett}, F.~H. and {Spencer}, R.~E. and {Paragi}, Z. and {Ogley}, R.~J. and {Cawthorne}, T.~V.},
        title = "{Radio-emitting component kinematics in SS433}",
      journal = {\mnras},
         year = {2002},
        month = {dec},
       volume = {337},
       number = {2},
        pages = {657-665},
          doi = {10.1046/j.1365-8711.2002.05944.x},
       adsurl = {https://ui.adsabs.harvard.edu/abs/2002MNRAS.337..657S}
}

@article{Punsly2021,
       author = {{Punsly}, B.},
        title = "{The Bulk Flow Velocity and Acceleration of the Inner Jet in M87}",
      journal = {\apj},
         year = {2021},
        month = {jul},
       volume = {918},
       number = {1},
        pages = {4},
          doi = {10.3847/1538-4357/ac0eee},
       adsurl = {https://ui.adsabs.harvard.edu/abs/2021ApJ...918....4P}
}

@article{Walker2018,
       author = {{Walker}, R.~C. and {Hardee}, P.~E. and {Davies}, F.~B. and {Ly}, C. and {Junor}, W.},
        title = "{The Structure and Dynamics of the Subparsec Jet in M87 Based on 50 VLBA Observations over 17 Years at 43 GHz}",
      journal = {\apj},
         year = 2018,
        month = mar,
       volume = {855},
       number = {2},
        pages = {128},
          doi = {10.3847/1538-4357/aaafcc},
       adsurl = {https://ui.adsabs.harvard.edu/abs/2018ApJ...855..128W}
}

@article{Biretta1999,
       author = {{Biretta}, J.~A. and {Sparks}, W.~B. and {Macchetto}, F.},
        title = "{Hubble Space Telescope Observations of Superluminal Motion in the M87 Jet}",
      journal = {\apj},
         year = {1999},
        month = {aug},
       volume = {520},
       number = {2},
        pages = {621-626},
          doi = {10.1086/307499},
       adsurl = {https://ui.adsabs.harvard.edu/abs/1999ApJ...520..621B}
}
\bibliographystyle{aasjournalv7}


\end{CJK*}
\end{document}